\documentstyle [prl,aps,epsf,floats]{revtex}

\def\eps{\epsilon}

\def\mh{m_{\rm h}}
\def\mw{M_{\rm W}}
\def\gw{g_{\rm w}}

\begin {document}

\def\OurTitlePage
{
\preprint {UW/PT-96-25}

\title  {
    Weakly first-order phase transitions:
    the $\epsilon$ expansion vs.\ numerical simulations
    \\in the cubic anisotropy model}.

\author {Peter Arnold, Stephen R.~Sharpe, Laurence G.~Yaffe, and Yan Zhang}

\address
    {%
    Department of Physics,
    University of Washington,
    Seattle, Washington 98195--1560
    }%
\date {\today}
\maketitle

\begin {abstract}%
{%
Some phase transitions of cosmological interest may be weakly first-order
and cannot be analyzed by a simple perturbative expansion around
mean field theory.
We propose a simple two-scalar model---the cubic anisotropy model---as
a foil for theoretical techniques to study such transitions, and we
review its similarities and
dissimilarities to the electroweak phase transition in the early
universe.
We present numerical Monte Carlo results for various discontinuities
across very weakly first-order
transitions in this model and, as an example, compare them to
$\eps$-expansion results.  For this purpose, we have computed through
next-to-next-to-leading order in $\eps$.

}%

\end {abstract}
}

\draft
\ifpreprintsty
  \OurTitlePage
  \newpage
\else
  \twocolumn[\hsize\textwidth\columnwidth\hsize\csname
    @twocolumnfalse\endcsname
    \OurTitlePage
  \vskip2pc]
\fi

Renormalization-group (RG) techniques for studying phase transitions,
such as the $\eps$ expansion,
have been used for over twenty years, and the analysis of second-order
transitions has produced quite successful quantitative predictions
\cite {Wilson,Zinn-Justin,cubic review}.
Weakly first-order transitions, in which the transition dynamics
is dominated by long but finite wavelength fluctuations,
were also investigated using $\eps$ expansion techniques
\cite {Rudnick,Domany,Chen}.
However, there has been less emphasis on testing quantitative predictions
for such first-order transitions,
in part because few predictions were easily accessible to
experiments of the time~\cite{footnote1}.

In recent years, renewed interest in quantitative predictions
for weakly first-order transitions has arisen in cosmology,
specifically the genesis of matter.
Current electroweak theory holds the promise of explaining
the net baryon number of the universe \cite{Kaplan}.
Any scenario of baryogenesis must satisfy Sakharov's conditions of
({\em i})
baryon number violation,
({\em ii})
C and CP violation, and
({\em iii})
departure from thermal equilibrium.
Electroweak theory
has all of these ingredients:
baryon number violation comes from a quantum anomaly,
C and CP violation is built in,
and a first-order electroweak phase transition can provide
the departure from equilibrium.
The electroweak transition separates the low-temperature world,
where the SU(2)$\times$U(1) gauge symmetry is hidden
and only the U(1) symmetry of electromagnetism is manifest,
from the high-temperature domain where the full SU(2)$\times$U(1)
symmetry is manifest.
To test the viability of electroweak baryogenesis,
detailed understanding of many properties of this phase transition
is required.

\begin {figure}
\vbox
    {%
    \vspace* {-17pt}
    \begin {center}
	\leavevmode
	
	\epsfbox {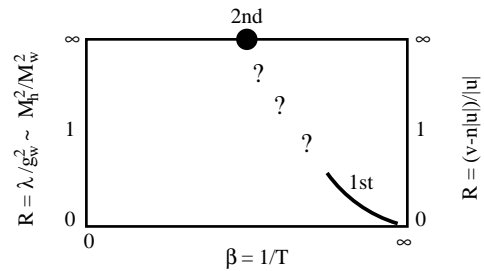}
    \end {center}
    \caption
	{%
	\label {fig:phases}
        Phase diagram of electroweak theory,
	or the cubic anisotropy model (for $u\le0, v\ge0$),
        based purely on analytic arguments.
	The left and right axes show the relevant ratio $R$ of couplings
	for electroweak theory, or for the cubic anisotropy model,
        respectively.
	}%
    }%
\end {figure}

Fig.~\ref{fig:phases}
sketches what can be firmly established about the phase diagram of
electroweak theory in the minimal Standard Model
(the standard model with a single SU(2) doublet Higgs field)
without resorting to numerical simulations.
A more complicated Higgs sector is probably needed for realistic
electroweak baryogenesis, but the minimal case is easier to discuss
and provides a useful testing ground.
The one important parameter of the minimal model not well determined
by experiment is the Higgs boson mass $\mh$,
or equivalently, the Higgs quartic self-coupling $\lambda$.
For a sufficiently light Higgs mass (or small $\lambda$)
the phase transition can be successfully treated by expanding
around mean field theory and is known to be fluctuation-induced
first order.
In other words, it appears to be second-order in mean field theory,
but perturbative fluctuations generate a finite correlation length
and render the transition first order.
For large Higgs mass (or large $\lambda$), the problem becomes
difficult to analyze.
Perturbative techniques are only reliable when
$\mh{\ll}\mw$, where $\mw \approx 80$ GeV is the W-boson mass.
In condensed matter applications,
this is commonly referred to as the Ginsburg criteria
for the success of mean field theory.
An equivalent way of stating this criteria is that the ratio of
couplings $R \equiv \lambda/\gw^2$,
where $\gw^2$ is the electroweak gauge coupling,
needs to be small.
If one (unrealistically) sends $\gw$ (and hence $\mw$) to zero,
so as to probe the opposite $R \to \infty$ extreme,
there is one piece of solid information.
The Higgs sector decouples from the gauge interactions in this limit,
and becomes a pure O(4) symmetric scalar theory,
which is known to have a second-order transition.

Viable electroweak baryogenesis generally needs as strong a first-order
phase transition as possible, and so prefers as small a Higgs mass as possible.
Experiment, however, puts lower bounds on Higgs masses
(60 GeV in the minimal model)
and forces one into the regime where a perturbative analysis of
the transition is at best marginal.
Consequently, there is a need to understand the full phase diagram
for electroweak theory and to develop techniques for reliable computations
even when $\mh \agt \mw$.

There are several things known about the transition in various unphysical
limits.  Near but just below 4 spatial dimensions ($d{=}4-\eps$),
the transition remains first order for
large Higgs mass \cite{Ginsparg,Arnold&Yaffe}, as we review below.
In the large $N$(scalar) limit (or in $2+\eps$ dimensions), there is a value
of $\mh$ at which there is a tricritical point, and the transition becomes
second-order for larger $\mh$
\cite{Arnold&Yaffe,March-Russel}.
In contrast, recent lattice simulations
suggest that the first-order line actually ends in a critical point
at $\mh \approx \mw$, above which there is no phase transition
whatsoever \cite{kajantie}.
There has been a recent attempt to understand this
behavior using RG methods in three dimensions \cite{Tetradis}.

In all the cases outlined above, one is left with the problem of how
to analyze the phase transition for those Higgs masses where it is
only weakly first-order and where perturbation theory breaks down.
For any techniques that purport to address this problem, it would be
useful to have an extremely simple statistical
system---on the level of simplicity of the Ising model---that could
serve as a canonical initial testing ground.
The Ising model provides a canonical example of a second-order transition
and is in the same universality class as the continuum field theory
of a single scalar field.
To study weak first-order transitions, one may add a second
spin field to make a model known as the Ashkin-Teller model, or
equivalently add a second scalar field to make a model known as
the cubic anisotropy model.

The purpose of the present work is to review these models; review
the similarities and dissimilarities with the electroweak phase
transition (see also \cite{Alford});
present the results of numerical simulations of
first-order transitions; and, as an example of testing a theoretical
method, compare the numerical results to predictions of the
$\eps$ expansion.
The $\eps$ expansion is the generalization of a theory from three to
$4{-}\eps$ spatial dimensions, where the dynamics of long wavelength
fluctuations is tractable using RG-improved perturbation theory,
followed by extrapolation of results back to $\eps{=}1$.
We have studied weakly
first-order transitions of the cubic anisotropy
model through next-to-next-to-leading
order in $\eps$.
(Leading order results were obtained twenty years ago in ref.~\cite{Rudnick}.)
We propose the cubic anisotropy model as a benchmark
for other investigators who advocate particular techniques for studying
weakly first-order transitions.
The details of our $\eps$ expansion calculations and numerical studies
are given elsewhere \cite{eps,numerical}.

  In its general form, the cubic anisotropy model is an O($n$) symmetric
scalar model to which an additional interaction is added which breaks O($n$)
down to hyper-cubic symmetry \cite{cubic review}:
\begin {equation}
   S = \! \int \! d^dx \left( {1\over2}\left|\partial\vec\phi\right|^2 
                 + {t\over2} \left|\vec\phi\right|^2
                 + {u\over4!} \left|\vec\phi\right|^4
                 + {v\over4!} \sum_{i=1}^n \phi_i^4
       \right) \!.
\label{eq:action}
\end {equation}
We shall focus on the simplest case, $n{=}2$.
This model is analogous to electroweak theory when $u\le0$ and $v\ge0$.
Fig.~\ref{fig:phases} illustrates the phase structure
for weak couplings ($|u|,v \ll 1$).
There is again a suitable ratio of couplings,
\begin {equation}
  R  \equiv  (v-n|u|)/|u| \,,
\end{equation}
which defines a Ginsburg criteria.
When the ratio drops below zero,
the tree-level potential in (\ref{eq:action}) is unstable,
just as $\lambda<0$ makes the Higgs potential unstable in electroweak theory.
When the ratio is small, perturbation theory turns out to be well behaved.
Then the phase transition is first order and, like electroweak theory, is
fluctuation driven.
When the ratio is moderate or large, perturbation theory breaks down.
When it is infinite, corresponding to $u{=}0$, the theory
reduces to $n$ decoupled copies of a theory of a single scalar field.
Each such copy is in the same universality class as the Ising model,
and has a second-order transition.

\begin {figure}
\vbox
    {%
    \vspace* {-40pt}
    \begin {center}
	\leavevmode
	
	\epsfbox {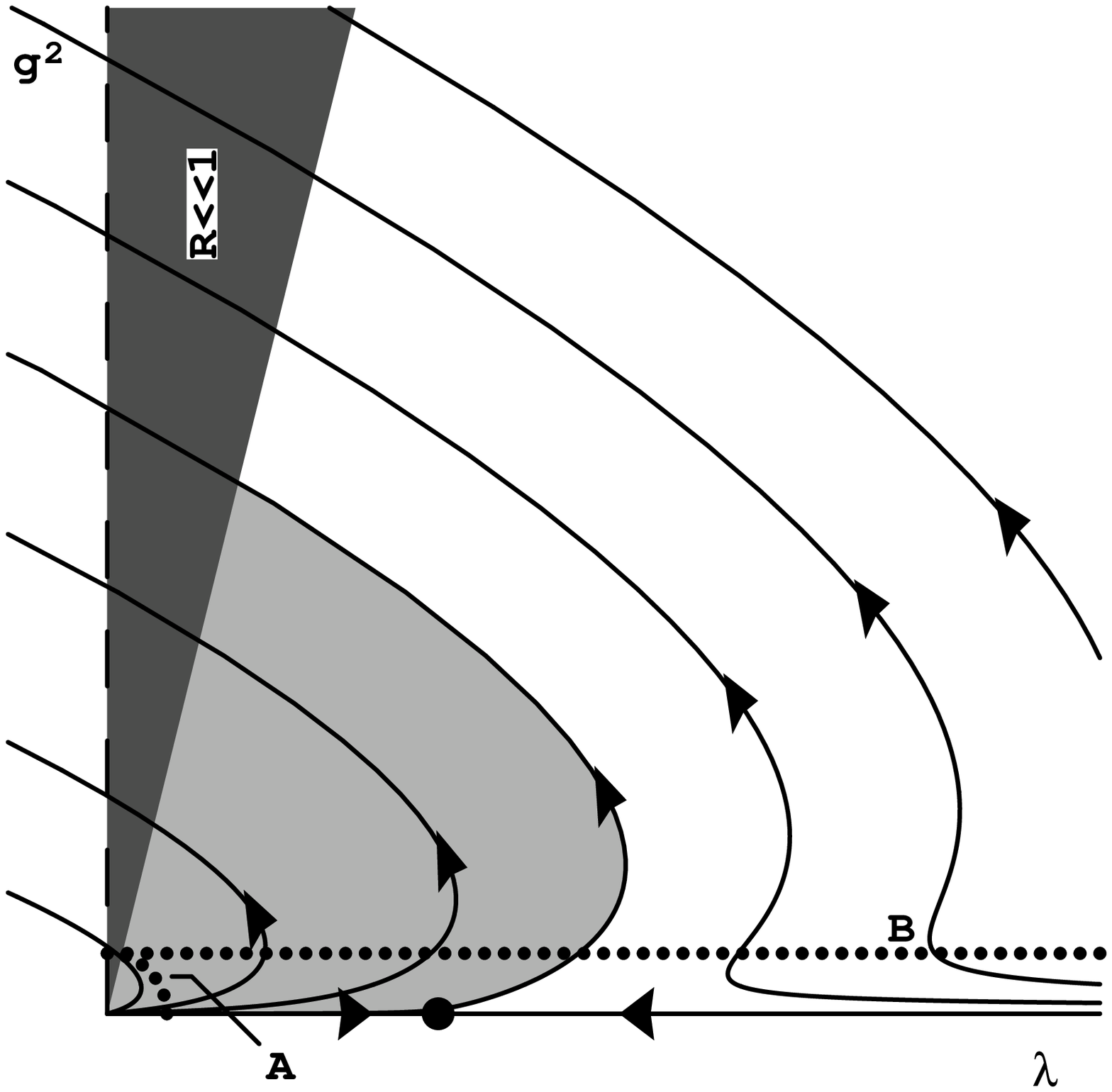}
	\vspace* {-60pt}
	\epsfbox {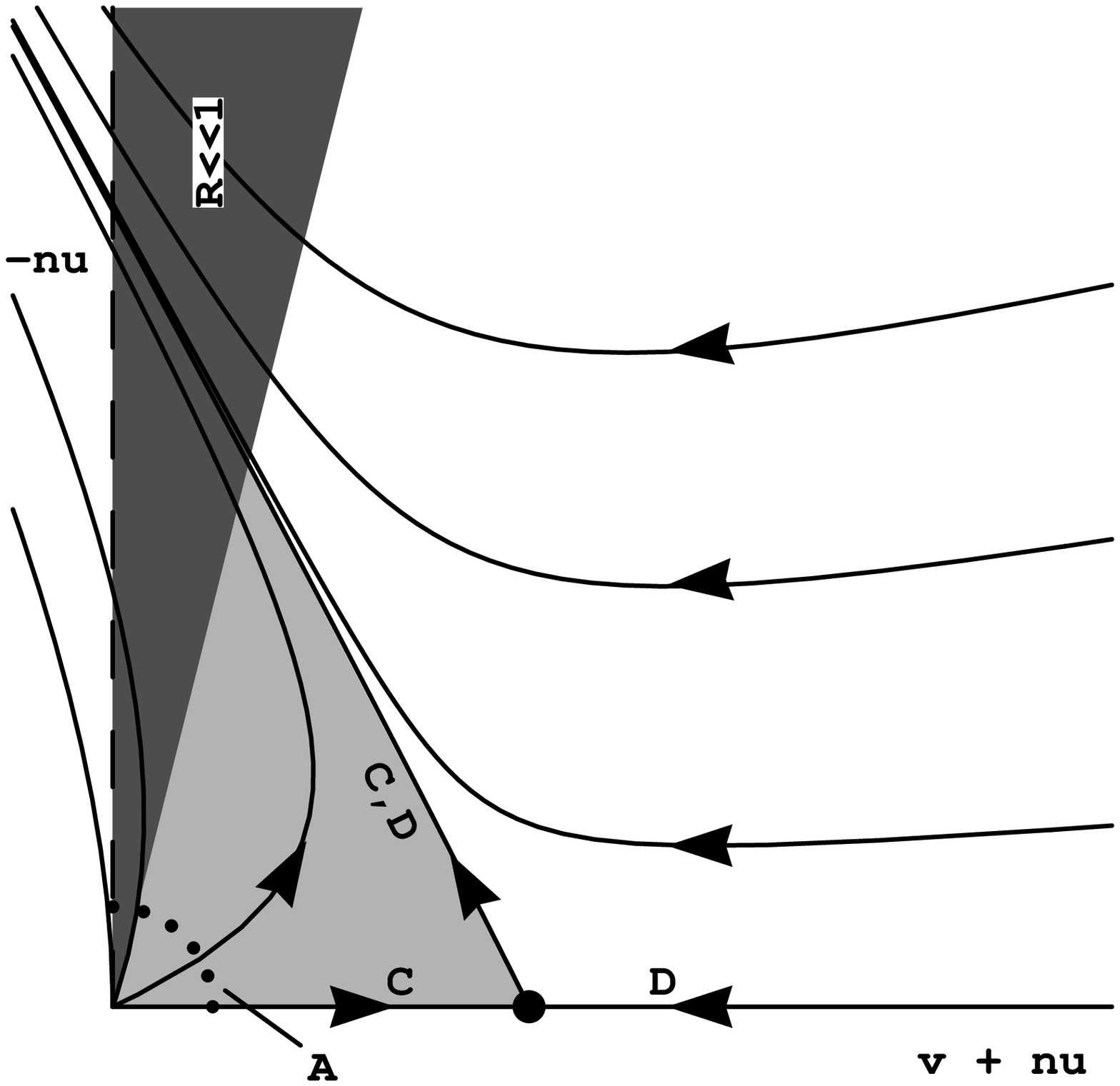}
    \end {center}
    \vspace* {-30pt}
    \caption
	{%
	\label {fig:flows}
        Renormalization group flows, for small $\eps$, of (a) electroweak
        theory, and (b) the cubic anisotropy model.
	}%
    }%
\end {figure}

To further appreciate the similarities between the cubic anisotropy model
and the electroweak phase transition, we turn to the $\eps$ expansion.
When $\eps$ is small, both theories can be solved using RG-improved
perturbation theory.
Fig.~\ref{fig:flows}a depicts the RG flow
of the couplings $\gw$ and $\lambda$
of electroweak theory in $4{-}\eps$ spatial dimensions
as one examines increasingly large distance scales.
Not shown is the third relevant parameter, the running Higgs mass,
which always grows (relative to the renormalization point)
as one scales to larger distances.
For $\gw{\not=}0$, the couplings do not flow to any fixed point.
They do, however, flow into the region $\lambda\ll \gw^2$, where the
Ginsburg criteria is satisfied.
After using the renormalization group to flow into this region,
one can use standard perturbative techniques to
make quantitative computations of thermal properties and
conclude that the transition is first order.
This is the argument that, for sufficiently small $\eps$,
the line of first order transitions shown in Fig.~\ref {fig:phases}
continues all the way to $R = \infty$.

   In condensed matter physics, the underlying theory at short distances
is typically strongly interacting.  In contrast, the Higgs model in
particle physics is generally of interest as a fundamentally {\it weakly}
coupled ($\lambda, \gw \ll 1$) theory at short distances
\cite{footnote2}.
Hence, one can focus attention on short-distance theories defined by
varying $\lambda/\gw^2$
in the neighborhood of the Gaussian fixed point at $\lambda{=}\gw{=}0$,
as illustrated by the arc $A$ near the origin in
Fig.~\ref{fig:flows}a.
By renormalization group equivalence, this covers all theories in the
lightly shaded region of fig.~\ref{fig:flows}a.
In particular, it covers the case
of the real world ($\gw$ fixed but $\lambda$ unknown, designated by line $B$),
provided the Higgs mass is not so large that the scalar sector is
fundamentally non-perturbative ({\it i.e.}, $\mh \alt$ 1~TeV).

Analogous flows for the cubic anisotropy model are show in
fig.~\ref{fig:flows}b.
For
$u\le0$ and $v\ge0$, the flows are
qualitatively similar to the electroweak case.
For $u=0$, the theory flows to the Ising fixed point.
For $u$ small and negative,
it instead flows to a region where the Ginsburg criteria is met and
the transition is known to be first order---an observation first exploited
by Rudnick \cite{Rudnick}.
The Ising fixed point is a tricritical point separating 
first-order ($u < 0$) and second-order ($u > 0$) transitions.

An important distinction between the cubic anisotropy model and
electroweak theory is that the cubic anisotropy model {\it must} have
a phase transition for any choice of the Ginsburg ratio $R$ because
a {\it global} symmetry is spontaneously broken.  No such argument
holds for electroweak theory
due to Elitzur's theorem \cite{Elitzur}.

In analogy with electroweak theory, one could attempt to analyze
the phase transitions in the cubic anisotropy model for weak couplings
$u$ and $v$,
and for arbitrary choices of the Ginsburg ratio $R$.
However, we know that the transition gets weaker,
and the analysis will become harder, as $R$ increases.
So an instructive goal is to analyze the extreme case
$R\to\infty$.
The limiting renormalization group trajectory is the one marked $C$ in
Fig.~\ref{fig:flows}b that runs from
the Gaussian fixed point to the Ising fixed point, and then away from the
Ising fixed point to the perturbative regime.
This extreme case has a substantial advantage for numerical simulations:
the same long-distance theory in the perturbative regime can be reached
starting from a {\it strongly} coupled short-distance theory
described by the
trajectory $D$ which comes in from ($u{=}0, v{=}\infty$) to the Ising fixed
point and then flows away to the perturbative regime along $C$.
The advantage of starting with this strongly coupled
theory is that it can be replaced by a discrete spin model,
which is easier to simulate numerically than theories
with continuous-valued fields.

In the (lattice regularized) theory of a single scalar field with
\begin {equation}
    S = \int d^d x \>
    \left[
	{\textstyle {1\over2}} (\partial\phi)^2 + \lambda(\phi^2-v^2)^2
    \right] \,,
\end {equation}
the field $\phi$ is constrained to only take on the two values $\pm v$ 
when $\lambda \to \infty$, and hence the theory reduces to a simple
Ising model with a discrete two component spin.
Similarly, the strong coupling limit of the $n=2$ cubic anisotropy model
is a model of two coupled Ising spins $s$ and $t$
known as the Ashkin-Teller model \cite{Ashkin,Ditzian}:
\begin {equation}
   S = - \beta \sum_{\langle ij \rangle}
       (s_i s_j + t_i t_j + x \, s_i s_j t_i t_j) \,,
\end {equation}
where the sum is over all neighboring pairs of sites $(i,j)$; the spins
$s,t={\pm} 1$; and $\beta$ and $x$ are parameters that
represent the inverse temperature and the deviation from the limit
of decoupled Ising models, respectively.
The $R \to \infty$ limit in our discussion of
the cubic anisotropy model corresponds to the $x{\to}0^+$ limit of the
Ashkin-Teller model.

   The $x{\to}0^+$ limit is crucial if one wants the Ashkin-Teller model
to correspond to the continuum cubic anisotropy model.
That is because generic first-order transitions have finite correlation
lengths of $O(1)$ lattice spacings at the transition.  Everything about
such transitions depends on the details of the short-distance
physics: the type of lattice used, the exact details of the couplings,
etc.  In the $x{\to}0^+$ limit, however, the correlation length at the
transition becomes arbitrarily large (since for $x=0$ it is infinite),
and so in this limit the details of the short-distance physics decouple.
Dimensionful quantities like the correlation length $\xi$ aren't directly
useful quantities to predict in this limit, since they diverge, but ratios
such as
\begin {equation}
   \lim_{x\to0^+} \> {\xi_+\over\xi_-} \equiv
   \lim_{x\to0^+} \> {\lim_{\beta\to\beta^-} \> \xi \over
                      \lim_{\beta\to\beta^+} \> \xi}
\label{eq:xiratio}
\end {equation}
are non-trivial and universal.  Here $\xi_+$ is the correlation length
infinitesimally above the critical temperature, and $\xi_-$ is that
infinitesimally below.
Our project has been to compute such ratios in the $\eps$ expansion for the
cubic anisotropy model and to compare to numerical simulations of the
corresponding
Ashkin-Teller model.

The details of our $\eps$ expansion calculations are given in ref.~\cite{eps}.
We find
\begin {eqnarray}
   {\chi_+\over\chi_-} &=&
   2 \, \bigl[ 1 + \eps \, (-0.1063\ln\eps + 0.4494) 
\\ &+&
      \eps^2(-0.0073\ln^2\eps
      + 0.1647\ln\eps - 0.2859)
      + O(\eps^3) \bigr] ,
\nonumber
\\
   {\xi_+^2\over\xi_-^2} &=&
   2 \, \bigl[ 1 + \eps \, (-0.1063\ln\eps + 0.4139) 
      + O(\eps^2) \bigr] ,
\\
   {C_+\over C_-} &=&
   \textstyle{17\over320} \, [ \, \eps + \eps^2(-0.2125\ln\eps + 1.3993)
      +O(\eps^3) ] .
\end {eqnarray}
$C$ is the specific heat.
$
    \chi \equiv \partial \langle \phi_1 \rangle / \partial h |_{h=0}
$
is the scalar susceptibility, defined
by adding an interaction $h\phi_1$ to the model,
and to leading order is the same as $\xi^2$.
Our leading-order results for the $\chi$ and $C$ ratios differ
by factors of 4 from those originally reported by Rudnick \cite{Rudnick}.

Table~\ref{table} summarizes the
leading-order and next-to-leading order results for our three
ratios.
In general, $\eps$ expansions are asymptotic expansions and the coefficients
begin to grow after just a few terms.
For critical exponents,
one can often get reasonably good values by
simply adding terms at $\eps{=}1$ until the contributions start to increase.
Our results in the cubic anisotropy model are not as well-behaved
as the best Ising model series, but are comparable to the $\eps$
expansion for the exponent $\eta$.
The expansion for $\chi_+/\chi_-$ leads one to expect that the actual
result should lie somewhere between the LO and NLO results.
One might have a similar expectation for $C_+/C_-$ and $\xi_+/\xi_-$,
even though the expansion for the former is clearly poorly behaved.

Very accurate predictions for critical exponents
in the Ising model have been obtained using
resummation techniques which combine knowledge of the asymptotic
large order behavior of the series with explicit low-order coefficients
\cite {Zinn-Justin}.
Our series expansions, however, contain logarithms of $\eps$,
and we do not currently know how to derive the large-order behavior
of these series.
As explained in ref.~\cite{eps}, these logarithms arise because physics
at the phase transition involves two scales whose ratio is $O(\eps)$.

Our numerical results are also presented in Table~\ref{table},
and details may be found in ref.~\cite{numerical}.
One sees that $C_+/C_-$ is indeed bracketed by the LO and NLO $\eps$-expansion
results and that $\xi_+/\xi_-$ works reasonably well.
$\chi_+/\chi_-$, however, is a disappointment for the $\eps$ expansion,
since the numerical result is not bracketed by LO and NLO, and the
the NNLO correction only makes the agreement worse.

Our results highlight the utility of the cubic anisotropy and Ashkin-Teller
models as foils for theoretical methods for making quantitative predictions
of weakly first-order transitions.  It would be interesting to see the
comparison of our numerical results against methods other than the
$\eps$ expansion, and it would be useful to have more accurate
Monte Carlo determinations of the universal ratios.

\begin{table}%
\begin {center}%
\tabcolsep=6pt
\begin {tabular}{c|lll|l}
universal ratio     & \multicolumn{3}{c|}{$\eps$ expansion}
                                      & Monte Carlo                   \\
                    & LO    & NLO   & NNLO &                          \\ \hline
$ C_+/C_-$          & 0.053 & 0.127 &      & 0.069(8)                 \\
$\xi_+/\xi_-$       & 1.4   & 1.7   &      & 1.7(2)                   \\
$\chi_+/\chi_-$     & 2.0   & 2.9   & 2.3  & 4.1(5)$^{\rm(a)}$        \\
\end {tabular}
\end {center}
\caption
    {%
    \label {table}
    Leading (LO), next-to-leading (NLO), and next-to-next-to-leading
    order (NNLO) results in the $\eps$ expansion
    compared to numerical Monte Carlo results.
    $^{(\rm a)}$ The error estimate on the Monte Carlo measurement of
    $\chi_+/\chi_-$ should be taken with a grain of salt
    (see ref.~\protect\cite{numerical} for details).
    }%
\end{table}

This work was supported by the U.S.~Department of Energy
grants DE-FG06-91ER40614 and DE-FG03-96ER40956.


\end {document}